# Calibration of higher eigenmodes of cantilevers

Aleksander Labuda[1], Marta Kocun[1], Martin Lysy[2], Tim Walsh[1], Jieh Meinhold[1],
Tania Proksch[1], Waiman Meinhold[1], Caleb Anderson[1] and Roger Proksch[1]

[1]*Asylum Research an Oxford Instruments Company, Santa Barbara, CA, 93117*
[2]*Department of Statistics and Actuarial Science, University of Waterloo, Ontario, Canada, N2L 3G1*

A method is presented for calibrating the higher eigenmodes (resonant modes) of atomic force microscopy cantilevers that can be performed prior to any tip-sample interaction. The method leverages recent efforts in accurately calibrating the first eigenmode by providing the higher-mode stiffness as a ratio to the first mode stiffness. A one-time calibration routine must be performed for every cantilever type to determine a power-law relationship between stiffness and frequency, which is then stored for future use on similar cantilevers. Then, future calibrations only require a measurement of the ratio of resonant frequencies and the stiffness of the first mode. This method is verified through stiffness measurements using three independent approaches: interferometric measurement, AC approach-curve calibration, and finite element analysis simulation. Power-law values for calibrating higher-mode stiffnesses are reported for several cantilever models. Once the higher-mode stiffnesses are known, the amplitude of each mode can also be calibrated from the thermal spectrum by application of the equipartition theorem.

## 1. Introduction

Atomic force microscopes (AFMs) use microscale cantilevers as transducers that convert forces between the nanoscale tip and sample into motion that can be measured with a macroscale photodetector. The accuracy in quantifying the nanoscale conservative and dissipative forces between the tip and sample is ultimately limited by the calibration of the cantilever stiffness and displacement measurements. This requirement has driven extensive research [1–22], and in the case of dynamic AFM techniques, where the cantilever is driven into oscillation, it has been focused almost exclusively on accurate determination of the stiffness and displacement sensitivity of the first cantilever eigenmode (resonant mode).

The recent rise in popularity of bimodal and multifrequency imaging[23–40], which provide high-resolution nanomechanical mapping of heterogeneous materials by exciting two or more cantilever eigenmodes, has extended the need for accurate cantilever calibration to its higher eigenmodes[41,42]. To date, the large uncertainty in higher-mode amplitudes and stiffnesses has impeded proper operation and quantitative data interpretation in multifrequency AFM. Uncertainty in these quantities has limited repeatability and accurate comparison to other measurement techniques.

This work demonstrates a rapid and simple method to calibrate the higher-eigenmode stiffnesses of cantilevers with arbitrary shapes. With calibrated cantilever stiffnesses, the sensitivity of the detection system for every eigenmode can also be calibrated through the equipartition theorem. This allows the determination of the amplitudes of every driven mode – all without touching a surface – thereby resolving the long-standing problem of uncertainty in stiffness and sensitivity during multifrequency imaging.

## 2. Overview: frequency-ratio calibration method

Recently, a calibration procedure[43,44] for the *first* eigenmode of cantilevers of arbitrary shapes was commercially implemented as GetReal™ by Asylum Research. Briefly, the idea is to meticulously characterize several reference cantilevers (of a particular type) with an interferometric measurement to precisely determine each eigenmode stiffness $k$, resonance frequency $f$, and quality factor $Q$ (see Appendix A(a) for details). Then, subsequent cantilevers of the same model can be calibrated in the field by measuring only their new $f$ and $Q$, and calculating their new stiffness via the scaling law $k \propto Qf^{1.3}$. Next, the equipartition theorem[45,46] is used to determine the optical beam deflection (OBD) sensitivity $S$ in units of nm/V (also called invOLS[19]). This calibration procedure, henceforth referred to as the "$Qf^{1.3}$ scaling" method, is visually described in Figure 1.

Although this calibration routine can in principle be applied directly to *higher* eigenmodes, the main limitation is that the thermal spectrum of higher modes has a considerably lower signal-to-noise ratio with respect to the first mode. Measuring a quality factor to a satisfactory precision may take several minutes, hours, or may even be impossible if the thermal response of the eigenmode is below the noise floor. On the other hand, measuring the resonance frequency of a higher mode from a thermal spectrum can be done very precisely and rapidly, as long as the thermal response is above the noise. Even in the absence of a thermal response, a driven measurement may provide an accurate estimate of the eigenmode resonance frequency, especially if photothermal[47–49] excitation is used.

Given the ease in measuring the higher-mode resonance frequency $f_n$ precisely, as opposed to quality factors[50], the higher-mode stiffness $k_n$ is commonly calibrated from the first-mode $f_1$ and $k_1$ by invoking the well-known scaling law

$$k_n = k_1 \left(\frac{f_n}{f_1}\right)^2 \quad (1)$$

(see Appendix B(a) for derivation). Unfortunately, this scaling law only applies to an ideal Euler-Bernoulli beam, which does not accurately describe real AFM cantilevers.

Real AFM cantilevers have a tip mass, and the tip is often positioned several microns from the cantilever end. Both of these effects cause an underestimation of $k_n$ when applying Eq. (1). Also, many common AFM cantilevers have a triangular section, which causes Eq. (1) to overestimate $k_n$. Analytical solutions to these three special cases of non-ideal cantilevers are presented in Appendix B(b,c,d) for reference. In practice, some combination of these competing geometrical effects, among others, contribute to deviation from Euler-Bernoulli behavior.

A simple modification to Eq. (1) that can largely account for these effects is to change the scaling law from square to some arbitrary power, as in

$$k_n = k_1 \left(\frac{f_n}{f_1}\right)^\zeta, \quad (2)$$

where the power-law exponent $\zeta$ can be determined empirically.

In practice, a representative sample of cantilevers from a particular type are carefully characterized *in factory* and used to estimate the power law exponent by

$$\zeta = \left\langle \frac{\log(k_n/k_1)}{\log(f_n/f_1)} \right\rangle, \quad (3)$$

where the brackets represent averaging over all cantilevers (see Appendix A(b) for derivation). Then, this power law exponent $\zeta$ can be used *in the field* to calibrate higher eigenmode stiffness $k_n$ via Eq. (2). This calibration procedure, henceforth referred to as the "frequency-ratio" method, is depicted in Figure 2.

Although the power-law model in Eq. (3) is not based on a fundamental physical principle, it provides a simple phenomenological description of the behavior of cantilevers that conveniently allows the estimation of $k_n$ from measured observables ($f_n, f_1, k_1$). Importantly, it avoids the use of the higher-mode quality factor $Q_n$. A more generalized model, where a different power law exponent $\zeta_n$ is used for each $n^{th}$ mode, may be more accurate for certain cantilevers, as will be investigated later.

The procedure for determining $\zeta$ for a particular cantilever type is the primary topic of this paper. Three methods will be compared: interferometric measurement, AC approach-curves, and finite element analysis simulation.

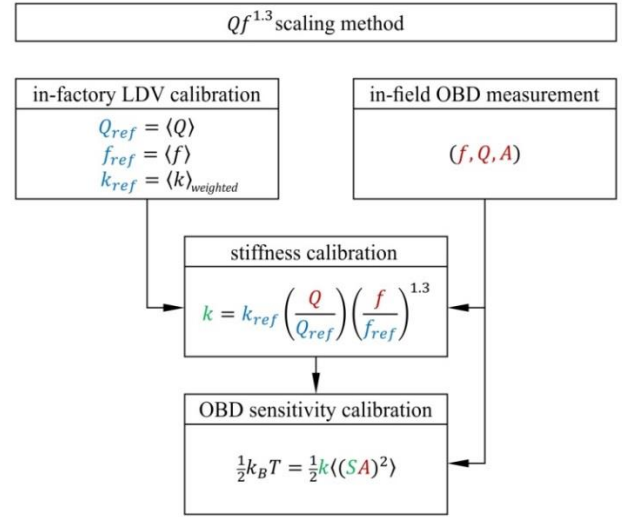

Figure 1: Graphical representation of the $Qf^{1.3}$ scaling method (implemented as GetReal™) for calibrating cantilever stiffness and OBD sensitivity. For a given cantilever model, a representative sample batch is selected and thoroughly characterized *in factory* with a laser Doppler vibrometer (LDV) measurement to obtain three reference parameters: $k_{ref}$, $Q_{ref}$, $f_{ref}$. The brackets $\langle \rangle$ represent averaging. Then, the stiffness $k$ for any cantilever of the same model is calibrated using measurements of $f$ and $Q$ in the field. Finally, the OBD sensitivity $S$, in units of nm/V, is obtained by satisfying the equipartition theorem through a measurement of the root-mean-squared amplitude $A$ of the cantilever thermal motion, in volts. The weighting of $\langle k \rangle$ is described mathematically in Appendix A(a).

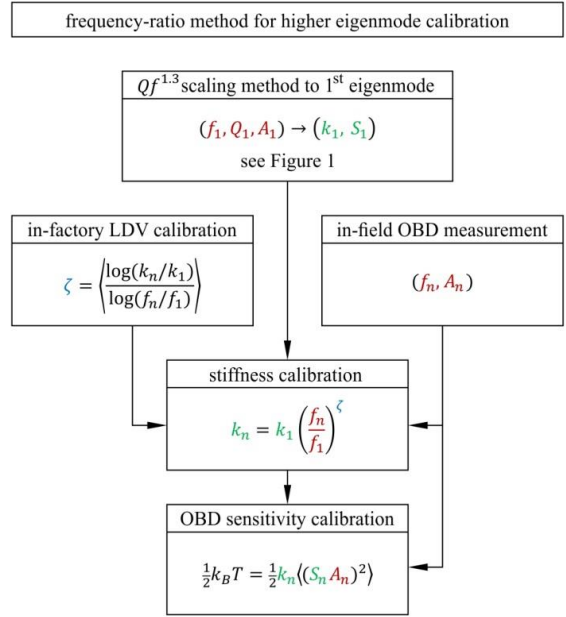

Figure 2: Graphical representation of the frequency-ratio calibration routine for the $n^{th}$ eigenmodes. The power law exponent $\zeta$ that relates $k_n/k_1$ to $f_n/f_1$ for a given cantilever model is characterized via a LDV measurement done *in factory* on a representative batch of reference cantilevers. After calibrating the first mode for a specific cantilever, the stiffness $k_n$ of that cantilever is calibrated from an additional measurement of the higher resonance frequency $f_n$. Finally, the eigenmode OBD sensitivity $S_n$, in units of nm/V, is obtained by satisfying the equipartition theorem through a measurement of the root-mean-squared amplitude $A_n$ of the cantilever thermal motion of the $n^{th}$ eigenmode, in volts.

## 3. Interferometric calibration

Perhaps the most direct way to calibrate the eigenmode stiffness is to measure its thermal motion interferometrically. The equipartition theorem relates the amplitude fluctuations to the stiffness as long as the cantilever motion is driven solely by thermal fluctuations of its environment, which is a very good approximation in ambient conditions.

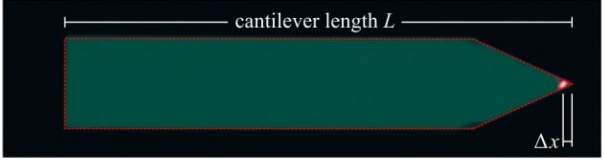

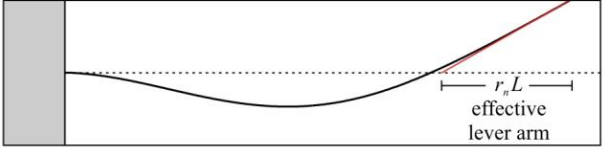

Figure 3: a) Photograph of an AC240 cantilever (Olympus) from the backside taken with the Cypher viewing optics. The cantilever length $L = 235.7$ μm. The LDV laser spot is located a distance $\Delta x = -4.0 \pm 0.7$ μm (to the left of the tip apex), leading to an underestimation of the true amplitude at the cantilever tip. The magnitude of the underestimation is calculable and depends on the eigenmode being measured. b) Schematic view of a mode bending shape. The effective lever arm $r_n L$ describes the slope of the cantilever end, as shown here for the second mode.

### a. Experimental setup

An Asylum Research Cypher AFM was retrofitted with a Polytec OFV-5000 laser Doppler vibrometer (LDV). The LDV's fiber-coupled laser ($\lambda = 633$ nm) was interfaced through the optical positioning system (a modified Cypher blueDrive[47] system retrofitted with broadband optics). This allows automated motion of the LDV laser spot (~2.5 μm diameter) with respect to the cantilever with sub-micron precision (see Ref. [51] for more details).

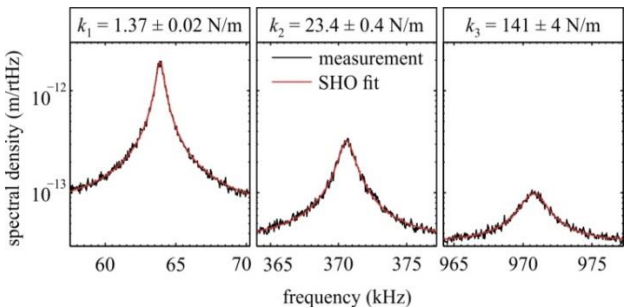

Figure 4: Spectral density of the first three eigenmodes of the AC240 cantilever from Figure 3. These data were calculated from a 2 s time series using the Daniell method[52] of PSD estimation to avoid bias and reduce effects of spurious noise peaks. Then, the spectra were corrected by the respective $\beta_n$ computed by Eq. (5). The quoted stiffnesses are the average of 37 consecutive such PSDs. The quoted errors were dominated by errors in $\beta_n$ due to the laser spot positioning uncertainty.

### b. Laser spot position correction

The main difficulty in calibrating several eigenmodes of a cantilever with an interferometric AFM stems from the need to accurately position the focused laser spot directly above the cantilever tip. For many cantilevers, the interferometric laser spot cannot be positioned exactly above the tip due to geometrical constraints. A power spectral density (PSD) measurement will underestimate the thermal noise at the cantilever tip if the laser spot is closer to the cantilever base, and vice versa. For small deviations $\Delta x$ between the laser spot and tip location, the correction factor to the amplitude spectral density (the square root of the PSD in units of m/$\sqrt{\text{Hz}}$) can be approximated as linear. Therefore, the amplitude measured at the spot location $A_\text{spot}$ relates to the amplitude at the cantilever tip $A_\text{tip}$ by

$$A_\text{tip} = \beta_n A_\text{spot}, \quad (4)$$

where the linear correction factor $\beta_n$ defined by

$$\beta_n = \left(1 + \frac{\Delta x}{r_n L}\right)^{-1}, \quad (5)$$

where $L$ is the length of the cantilever and $r_n$ is the ratio that determines the effective lever arm length of the $n^\text{th}$ mode, as graphically represented in Figure 3. Formally, the effective lever arm $r_n L$ is the distance between the cantilever end location (on the x-axis) and the location where the linear fit to the cantilever end intersects with the x-axis.

Note that if the laser spot is closer to the base relative to the tip, $\Delta x$ is negative and $\beta_n > 1$. Expressing the correction factor in terms of the ratio $r_n$ is convenient since $r_n$ is a very weak function of the cantilever length $L$ and does not change appreciably for different cantilevers of a particular model. As discussed in the next section, once $r_n$ is determined for a specific cantilever type, $L$ can be easily measured for every single cantilever as shown in Figure 3 and used to correct the measured spectral density by the appropriate $\beta_n$ factor.

Although many cantilever types have a tip setback[1] that allows the laser spot to be positioned immediately above the tip, the $\beta$ factor is still used for error analysis, especially at higher eigenmodes.

### c. Stiffness measurement

The stiffnesses of the first three eigenmodes of the cantilever presented in Figure 3 were measured using the methodology described above. The results are shown in Figure 4. A measurement of the local slope of each mode at the cantilever tip location is required to define $r_n$ for the correction in Eq. (5). Each $r_n$ can be obtained by locally fitting the $n^\text{th}$ mode cantilever shape, obtained by the shape mapping method presented in the next section. The uncertainty in stiffness estimation was dominated by errors in laser spot positioning (relative to the tip) and was calculated by Eq. () in Appendix A(c).

---

[1] Tip setback is defined as the distance between the cantilever end and the location of the tip (measured in the cantilever plane).

### 4. Calibration through AC approach curves

The optical beam deflection (OBD)[53,54] system used to measure cantilever deflection in nearly all AFMs has the disadvantage of lacking an absolute calibration reference. Additionally, the OBD sensitivity depends on the mode shape of the cantilever[55,56] and requires separate calibration of each mode to determine the corresponding mode stiffness, as presented in this section.

#### a. Sensitivity calibration

The method used here for calibrating the OBD sensitivity for a specific eigenmode starts by approaching a stiff sample while driving the cantilever on resonance at the desired mode with a large amplitude. In high-$Q$ environments, the amplitude of the cantilever decreases symmetrically[57] and approximately linearly with respect to tip-sample distance[58]. This implies that the reduction in cantilever amplitude (in volts) can be calibrated against the approach of the sample towards the cantilever (in nanometers) to obtain an estimate of the OBD sensitivity for the driven eigenmode. Example AC approach curves for three eigenmodes of a cantileverare presented in Appendix A(d).

#### b. Stiffness measurement

With OBD sensitivities $S_n$ for the n$^{\text{th}}$ eigenmode calibrated in the previous subsection, the equipartition theorem may be applied to extract the respective eigenmode stiffnesses from the thermal PSD[45,46] (not shown) by

$$k_n = \frac{k_B T}{\langle (S_n A_n)^2 \rangle}, \quad (6)$$

where $A_n$ is the root-mean-squared amplitude of the cantilever thermal motion of the $n^{\text{th}}$ eigenmode (in volts), $k_B$ is the Boltzmann constant, and $T$ is temperature.

#### c. Advantages and limitations

The advantage of this technique over interferometric measurements is that the OBD laser spot position with respect to the tip requires no correction. The drawback, however, is that tip-sample interaction nonlinearities and instabilities can easily degrade the accuracy of the calibration[46,59]. In this study, several approach curves were performed and those that exhibited the most linear behavior were selected for analysis.

For this method, error analysis was not performed. The largest sources of uncertainty are the subjective choices for the fitting range of the selected AC approach curves, as well as the assumptions that the decrease in tip-sample distance maps directly to an amplitude decrease. These problems, along with the potential tip contamination or damage caused by this method, motivate the other calibration approaches outlined in this work.

Finally, a noteworthy limitation of the AC-approach calibration of soft cantilevers (roughly $k_1 < 1$ N/m) was unsuccessful because the snap-to-contact of the first mode prevented stable approach curves while driving higher modes.

### 5. Calibration through FEA simulation

Finite element analysis (FEA) simulations[60] were performed to provide an independent measurement of the power law relating stiffness to frequency of higher eigenmodes. The validation of these FEA simulations was done by comparing the simulated FEA mode shapes to measured LDV mode shapes.

#### a. LDV mode shape mapping

While the cantilever is piezoacoustically driven at one of its eigenmodes, the LDV spot is translated along the cantilever axis and the amplitude $|A_n|$ and phase $\phi_n$ at every location $x$ are measured by a lock-in amplifier, where $x$ represents the distance along the cantilever normalized by its full length $L$. The eigenmode shape for the $n^{\text{th}}$ mode is then reconstructed as

$$\psi_n(x) = |A_n(x)| \cos \phi_n(x), \quad (7)$$

where the phase at the cantilever end $\phi_n(x=1) \equiv 90°$. Not only does the phase correction unwrap the shape at higher modes, it also removes any cantilever base motion from the measurement caused by piezoacoustic excitation, which is out of phase with the cantilever end motion[61,62] and irrelevant to the analysis in this study. Finally, each mode shape is normalized to enforce $\psi_n = 1$ at the cantilever tip location.

This protocol was used to map the first three eigenmodes of an AC240 cantilever, which are presented in Figure 5 and were used to extract the value of $r_n$ for each mode.

#### b. FEA simulation

Finite element analysis (FEA) simulations were performed using SolidWorks (Dassault Systèmes, Waltham, MA). The plan-view dimensions of the modeled cantilever were taken from photographs such as those in Figure 3. The cantilever chip was also modeled to extend several tens of micrometers beyond and around the cantilever base to account for the chip's finite stiffness[63]. The need for modeling the cantilever chip was assessed in a benchmark experiment presented in Appendix A(e).

The thickness of the modeled cantilever was tuned such that the first-mode FEA resonance frequency matched the LDV measured value. Additionally, the cantilever model included a slight taper with linearly varying thickness along the cantilever length (thinner at the end). Adjusting the taper and the tip height (within the manufacturer's tolerance) provided better agreement between FEA simulations and LDV measurements. The justification for these adjustments is that the taper and tip height combination reduced the discrepancy between the FEA and LDV mode shapes as well as simultaneously reduced differences between the FEA and LDV resonance frequencies of all three eigenmodes. A taper in the thickness of the FEA model may also effectively account for a variation in the stiffness of the real cantilever, either due to a gradient in the Young's modulus[64] or to gradients in the surface stress that affect the local cantilever stiffness[65].

### c. Stiffness calculation

Next, the stiffness was calculated from the FEA eigenmode shapes by the following integral[66]:

$$k_n = \int_{-\infty}^{1} \frac{EI(x)}{L^3}\left(\frac{d^2\psi_n(x)}{dx^2}\right)^2 dx, \quad (8)$$

where $E$ is Young's modulus, and $I(x)$ is the second moment of area. The $-\infty$ integration limit in Eq. (8) indicates the need to start the integration inside the cantilever chip, rather than at the cantilever base at $x = 0$, due to the non-negligible deformation of the cantilever chip described earlier.

## 6. Results and discussion

In this section, the validity of the power-law model to relate eigenmode stiffnesses to resonance frequencies will first be assessed, followed by a detailed error analysis that quantifies the model's predictive power. The merits of this calibration procedure will be discussed and compared to other calibration models.

### a. Power-law model validity assessment

The data acquired by LDV measurements, AC approach curves, and FEA simulations are plotted together in Figure 6. Also shown is a power-law fit to the second-mode LDV measurements. Very good agreement in stiffness (on the order of 10%) between all three methods and the power-law fit (for the second and third modes) provides confidence that the model proposed by Eq. (2) is appropriate for the second and third eigenmodes of an AC240. Furthermore, the FEA data for the fourth and fifth modes (not shown) suggests that the power law extends to higher eigenmodes with a reasonable degree of accuracy.

As mentioned earlier, there is no physical basis for relating all the stiffness ratios to frequency ratios by the *same* power law, and it is more accurate to fit a different power law exponent $\zeta_n$ for each $n^{th}$ mode, if enough data is available. Given the existence of LDV data at both the second and third modes, it is worthwhile analyzing them separately to obtain separate power-law exponents $\zeta_2$ and $\zeta_3$ to improve the accuracy of each eigenmode model. For the data in Figure 6, $\zeta_2 = 1.72(\sigma = 0.01)$ and $\zeta_3 = 1.68 (\sigma = 0.01)$, where $\sigma$ represents the standard deviation of all measured $\zeta_n$ values for this batch of cantilevers. Although the exponents are distinguishable within error, assuming an average value $\zeta = (\zeta_3 + \zeta_2)/2$ in this example would have led to an underestimation of only 3% in $k_2$ and an overestimation of only 5% in $k_3$. Nonetheless, in the presence of empirical data, separate exponents $\zeta_n$ for each $n^{th}$ mode provide higher accuracy.

Next, the validity of the power-law model is assessed across different cantilever types, with reference to the data summarized in Table 1 and scanning electron microscope images in Figure 7. As predicted by the analytical modeling in Appendix B, the rectangular cantilevers with a tip setback indeed have $\zeta > 2$, while the cantilevers with distinctively triangular tips result in $\zeta < 2$. Furthermore, this qualitative trend agrees with all measurable power-law exponents $\zeta_n$ of higher modes. This suggests that the near power-law behavior stems from real and predictable geometrical effects.

$\zeta = 2$ is often assumed when estimating higher eigenmode stiffnesses (e.g., see Eq. (1)). This assumption can be far from valid, and performing an empirical measurement of $\zeta$ leads to a considerable gain in accuracy. While the $\zeta = 2$ assumption underestimates $k_2/k_1$ by $\sim 1.6 \times$ for an AC240 cantilever, using the LDV-measured value $\zeta = 1.72$ results in an estimate of $k_2/k_1$ accurate to a few percent.

Because the power-law model is only a phenomenological description of cantilever eigenmode properties that combines several geometrical effects, empirical justification of its use is always required. Ideally, $\zeta_n$ for every eigenmode is measured. However, assuming that $\zeta_{n+1} = \zeta_n$ in the absence of an empirical measurement of $\zeta_{n+1}$ may be reasonable and is likely much more accurate than assuming $\zeta = 2$. This is backed by analytical predictions of near power-law behavior for various geometrical effects, FEA simulations of real cantilever geometries, as well as LDV measurements on various cantilevers.

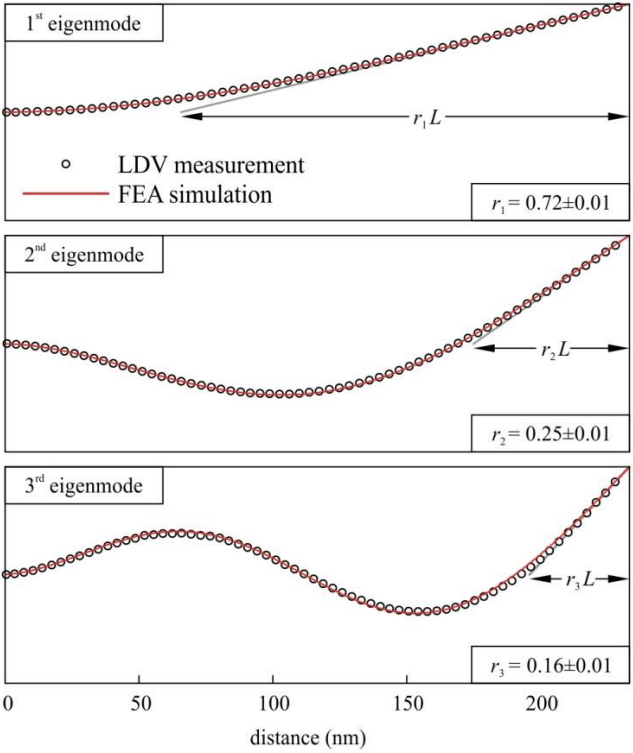

Figure 5: LDV measurements and FEA simulations of the first three eigenmodes of an AC240 cantilever. For the FEA simulations, the cantilever thickness was tapered from 2.58 μm at the base to 2.28 μm at the tip, whose height was taken to be 13.0 μm. Both are within the manufacturer's tolerance range. The values of $r_n$ for each lever are also shown. Note that the deflection of these driven modes (along the vertical axis) is highly exaggerated for visual reasons; the true deflections never exceeded 100nm.

### b. Power law model error analysis

In Figure 6, the random errors on the LDV measurements are much smaller than the deviations from the power-law fit, which can be attributed to epistemic error. Epistemic error represents the inability of a model to capture all the physics of some physical system. In this case, the power-law model cannot describe all the variability in this batch of AC240 cantilevers. In fact, there is no reason to believe that all micromachining variability leads specifically to power-law behavior; therefore, such an assumption will always lead to some calibration uncertainty.

The standard deviation of the power-law exponent $\sigma_{\zeta_n}$ measured for a population of cantilevers can be used in future calibrations to define an error in stiffness ratio by

$$\sigma_{k_n/k_1} = \sigma_{\zeta_n} \left(\log \frac{f_n}{f_1}\right)\left(\frac{f_n}{f_1}\right)^\zeta . \quad (9)$$

This relationship was derived under the assumption that errors in stiffness are multiplicative and that relative errors are small. (See Appendix A(b) for derivation)

Errors in measuring $f_n$ and $f_1$ can be safely ignored in typical experimental settings, as they will be dominated by $\sigma_{\zeta_n}$, which was $\sigma_{\zeta_2} = 0.008$ for the second eigenmode data in Figure 6. Since $\sigma_{\zeta_2}$ is dominated by the epistemic model errors rather than measurement errors, averaging over more measurements of stiffness would not reduce $\sigma_{\zeta_2}$. Note that the error bars in Figure 6, representing random measurement error, are much smaller than the scatter of the data around the power law fit line.

In this study, the measured standard deviation $\sigma_{\zeta_2} = 0.008$ corresponds to a stiffness-ratio error $\sigma_{k_2/k_1} = 1.5\%$. For the third mode, the standard deviation $\sigma_{\zeta_3} = 0.014$ correspond to a stiffness-ratio error $\sigma_{k_3/k_1} = 3.8\%$. These low errors demonstrate the efficacy of this calibration method, which is particularly accurate because it is based on stiffness *ratios*. Systematic calibration errors in the measurement drops out (e.g., LDV sensitivity) when dividing $k_n$ by $k_1$. In fact, these errors in stiffness ratios may be lower than the calibration error in determining the first-mode stiffness $k_1$, which relies on *absolute* accuracy.

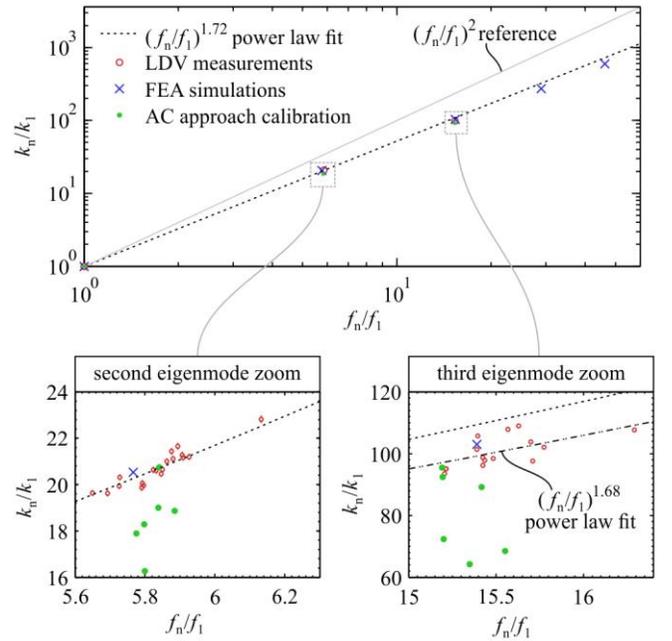

Figure 6: Stiffness ratio $k_n/k_1$ versus frequency ratio $f_n/f_1$ for the first three eigenmodes of an AC240 cantilever as measured with three different methods. The LDV measurements were performed on 19 cantilevers taken from 11 different microfabrication wafers. Six of those cantilevers were used to perform AC approach calibration measurements. The second mode LDV data were used to generate the power-law fit with exponent $\zeta_2 = 1.72$ shown on all the graphs. In the bottom right graph, a power-law fit with $\zeta_3 = 1.68$ specifically for the third mode data is also shown. The errors bars on the LDV data points represent random errors (standard deviation) as deduced from repeated measurements on the same cantilevers; the error bars for the third eigenmode cannot be seen because they are smaller than the data points.

| Cantilever model | $\zeta_2$ | $\zeta_3$ | $\zeta_4$ | $f_1$ (kHz) | $k_1$ (N/m) | M | W |
|---|---|---|---|---|---|---|---|
| AC240 | 1.72(0.01) | 1.68(0.01) | **N/A** | 70 | 2 | 19 | 11 |
| AC200 | 1.67(0.01) | **N/A** | **N/A** | 115 | 10 | 11 | 5 |
| AC160 | 1.67(0.02) | **N/A** | **N/A** | 300 | 26 | 14 | 8 |
| Arrow-CONT | 1.94(0.01) | 1.93(0.01) | 1.91(0.01) | 14 | 0.2 | 16 | 4 |
| CONT | 2.09(0.02) | 2.11(0.03) | 2.14(0.03) | 13 | 0.2 | 42 | 13 |
| FM | 2.13(0.05) | 2.17(0.07) | **N/A** | 75 | 2.8 | 22 | 12 |
| NCH | 2.17(0.06) | **N/A** | **N/A** | 320 | 42 | 17 | 10 |

Table 1: Values of the power-law exponent $\zeta_n$ measured by LDV for the $n^{th}$ eigenmode for different cantilever types. Nominal values of $f_1$ and $k_1$ are also shown for reference. M is the number of cantilevers per type; W is the number of microfabrication wafers that were sampled. Numbers in brackets are the standard deviation of the epistemic errors of the power-law model. They dominate over other sources of measurement variability (see Appendix A(c) for details about error analysis). For certain cantilever types, higher modes could not be measured due to LDV bandwidth and noise limitations.

From this batch of AC240 cantilevers, the epistemic error in measuring $k_1$ using the $Qf^{1.3}$ scaling method resulted in a normalized error $\sigma_{k_1} = 5.3\%$. This error stems from inaccuracies of the $k \propto Qf^{1.3}$ scaling-law model in describing the cantilever parameter space that leads to stiffness variability. Because $\sigma_{k_1} > \sigma_{k_2/k_1}$, the error in $k_2$ of an AC240 is actually dominated by the error in measuring its $k_1$ and not significantly affected by the power-law error $\sigma_{\zeta_2}$ introduced by the use of Eq. (2). In other words, errors in calibrating $k_1$ are the limiting factor in calibrating $k_2$ for this cantilever type.

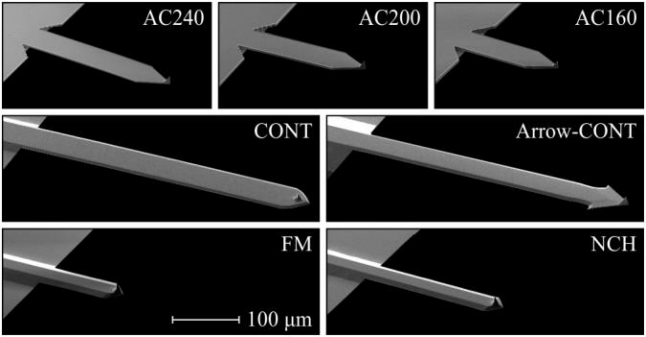

Figure 7: SEM photographs of the cantilevers presented in Table 1. The scale bar relates to all images. These images were photoshopped to remove distracting elements in the background.

c. **Representative sampling**

To ensure that a representative sampling of cantilevers was selected for each cantilever model, the cantilevers were selected from as many microfabrication wafers as available. Furthermore, a variety of coatings were selected when possible; for example, the FM category contains some cantilevers that are uncoated, and others that are coated with Au, Al, PtIr, and PtSi. Also, some models were obtained from NanoWorld and others from Nanosensors.

Representative sampling of cantilevers enables drawing conclusions about the micromachining variability across different wafers. The AC160, AC200, AC240, ArrowCONT are particularly reproducible due to their fabrication process that requires no alignment between the tip and the cantilever; this avoids variability in tip setback. These models have a distinctively smaller epistemic error with respect to the CONT and FM models, which have the disadvantage of having a tip setback that introduces variability in stiffness, especially at higher eigenmodes.

d. **Comparison of calibration models**

Even in the absence of interferometric detection, it may be more accurate to calibrate higher eigenmodes with AC approach curves to provide a measured value of $\zeta$ rather than assuming $\zeta = 2$. This can be concluded from Table 2, where the LDV and AC approach-curve measurements show reasonable agreement. Unfortunately, as mentioned earlier, soft cantilevers could not be calibrated successfully using AC approach-curves due to snap-to-contact of the first mode while driving higher eigenmodes.

It is also worth assessing the frequency-ratio calibration method with respect to simply applying the $Qf^{1.3}$ scaling method to higher modes independently.

Table 3 summarizes the epistemic errors for both cases. Since epistemic errors refer specifically to the inability of each model to predict the true stiffness, they do *not* consider random measurement errors that were made insignificant by long averaging of thermal noise, nor do they include systematic calibration errors. Note that both models were applied to the same dataset of meticulously calibrated cantilevers. Also, the dataset was measured twice (two weeks apart) and resulted in nearly identical errors. The conclusion is that calculating $k_2$ and $k_3$ by frequency-ratio calibration is more accurate by roughly a factor of 2 than calibrating by the use of the $Qf^{1.3}$ scaling method for each higher eigenmode independently. A similar conclusion was drawn for all the cantilever types studied in the context of this work. The lower error of the frequency-ratio method is attributed to the fact that the $Q$ factor of the higher mode is omitted from the measurement. Not only does omitting the $Q$ factor measurement increase accuracy, it increases the precision in estimating $k_2$, because resonance frequencies are easy to measure precisely. Omitting the $Q$ factor measurement, which is prone to bias and measurement error, also has the benefit of making the frequency-ratio calibration method much more robust.

Table 3 also compares the epistemic errors that arise in estimating stiffness ratios $k_n/k_1$ directly for the AC240 cantilever. The error in this ratio is relevant to calibration for bimodal imaging, because the relative change in stiffness of two driven eigenmodes is used as a metric[27]. Notably, the frequency-ratio method outperforms the $Qf^{1.3}$ scaling method in this context as well. This improvement is attributed to the fact that the error-prone quality factor $Q_n$ of higher modes is omitted in the frequency-ratio method.

| cantilever | $\zeta_2$ | |
|---|---|---|
| | LDV | AC approach |
| AC240 | 1.72 (σ = 0.01) | 1.66 ± 0.02 |
| AC200 | 1.67 (σ = 0.01) | 1.69 ± 0.03 |
| AC160 | 1.67 (σ = 0.02) | 1.62 ± 0.02 |
| FM | 2.13 (σ = 0.05) | 2.22 ± 0.06 |

Table 2: Power-law exponent for the second eigenmode $\zeta_2$ for different cantilevers obtained from LDV and AC approach curve measurements. The standard deviations (σ) of LDV-measured $\zeta$ values are dominated by epistemic (model) error. The errors in the AC approach data are the standard error (±) of the mean calculated from several approach curves.

| Estimated parameter | $Qf^{1.3}$ scaling method | Frequency-ratio method |
|---|---|---|
| $k_1$ | 5.3% | N/A |
| $k_2$ | 9.7% | 5.4% |
| $k_3$ | 14.4% | 5.7% |
| $k_2/k_1$ | 3.1% | 1.5% |
| $k_3/k_1$ | 4.1% | 3.8% |

Table 3: Epistemic (model) error for the $k \propto Qf^{1.3}$ scaling method and the frequency-ratio method on eigenmode stiffnesses and their ratios for a batch of 19 AC240 cantilevers sampled from 11 different wafers. Note that all random measurement errors were shown to be insignificant by acquiring two independent datasets; both datasets resulted in nearly identical percentages.

**e. Absolute accuracy and bandwidth limitations**

Absolute accuracy was not discussed so far in this analysis of the frequency-ratio method. This allowed for an assessment of the quality of the power-law model used in Eq. (2), while disregarding any absolute accuracy errors in calibrating $k_1$. Although the *absolute* accuracy of all eigenmode stiffnesses is fundamentally limited by the accuracy in the LDV calibration itself, such calibration errors do not affect the accuracy of stiffness *ratios*.

However, any frequency dependence of the LDV response directly translates into accuracy errors in stiffness ratios. The magnitude of the error for this frequency-dependence was estimated by measuring the eigenmodes of a tipless cantilever that closely resembles an ideal Euler-Bernoulli beam (see Appendix Ae). It was concluded that the frequency-dependence of the LDV is small compared to the epistemic errors for the data presented in this paper.

The finite bandwidth (2.5 MHz) and detection noise (~15 fm/$\sqrt{\text{Hz}}$ at high frequency) of the LDV prevented the accurate acquisition of the $\zeta_n$'s that are missing in Table 1.

**f. Sensitivity calibration**

An important direct benefit derived from having calibrated eigenmode stiffnesses is that the OBD sensitivity can be determined from a PSD for every mode without contacting the sample. This leads to an accurate measure of the amplitude of higher modes, which typically remains unknown in multifrequency AFM experiments. The non-invasive calibration of higher mode sensitivities proposed here can be used to standardize protocols for multifrequency AFM experiments and provide more meaningful comparison between different experiments.

## 7. Conclusion

A semi-empirical power law model was proposed for calibrating higher eigenmodes of cantilever of arbitrary shape with minimal effort on the part of the AFM user. The outlined procedure calibrates the stiffness of *higher* modes with respect to *first* mode stiffness, thereby leveraging efforts invested in calibrating cantilevers in previous studies. Once a particular cantilever type is characterized in factory with a power-law exponent $\zeta$, only the resonant frequencies of eigenmodes are necessary for calibrating higher mode stiffnesses prior to an AFM experiment. By avoiding the need for the AFM user to perform difficult measurements of higher-mode $Q$ factors or detection sensitivity, the calibration procedure provides rapid and accurate results in experimental settings. With calibrated eigenmode stiffnesses, the detection sensitivity of higher modes can also be obtained before ever contacting the sample.

These benefits translate directly to quantitative bimodal and multifrequency AFM techniques that rely on accurate eigenmode stiffnesses to provide accurate nanomechanical properties of the sample, as well as an accurate measure of the amplitudes of higher modes that affect the interpretation of tip-sample interaction physics.

## 8. Acknowledgements

The authors acknowledge careful correction of the manuscript by Donna Hurley, and useful discussions with Deron Walters, Jason Cleveland, Mario Viani, and Teimour Maleki.

## 9. Appendix A

### a. Population averaging for $Qf^{1.3}$ scaling method

The stiffness $k$ of an uncalibrated cantilever in the field can be determined from its resonance frequency $f$ and quality factor $Q$ via

$$k = k_{\text{ref}} \left(\frac{Q}{Q_{\text{ref}}}\right) \left(\frac{f}{f_{\text{ref}}}\right)^{1.3}, \tag{A1}$$

where the reference parameters are measured in factory by the manufacturer from a representative sample of $N$ test cantilevers. The reference parameters can be computed by

$$f_{\text{ref}} = \langle f \rangle, \tag{A2}$$
$$Q_{\text{ref}} = \langle Q \rangle, \tag{A3}$$

and

$$k_{\text{ref}} = \langle k \left(\frac{Q_{\text{ref}}}{Q}\right) \left(\frac{f_{\text{ref}}}{f}\right)^{1.3} \rangle, \tag{A4}$$

where the brackets represent averaging over the representative sample of $N$ reference cantilevers. Note that more elaborate averaging schemes exist[44] that lead to different $k_{\text{ref}}$, $Q_{\text{ref}}$, and $f_{\text{ref}}$ but result in identical estimation of $k$.

### b. Population averaging for frequency-ratio method

Defining the frequency ratio $\tilde{f} = f_n/f_1$ and stiffness ratio $\tilde{k} = k_n/k_1$ simplifies Eq. (2) into

$$\tilde{k} = \tilde{f}^{\zeta}. \tag{A5}$$

Each cantilever from a population has an exact value

$$\zeta^* = \frac{\log \tilde{k}}{\log \tilde{f}}, \tag{A6}$$

that may differ from the population average $\zeta$. The epistemic (model) error, which was referred to throughout the text, can be treated by assuming the distribution of all $\zeta^*$ values from a population has a variance $\sigma_\zeta^2$ which inevitably leads to some error when estimating $\tilde{k}$ of any particular cantilever based on a measurement of $\tilde{f}$. This error stems from the assumption that a single $\zeta$ applies to all cantilevers from the population, and can be represented by an error term $\epsilon$, as in

$$\tilde{k} = \tilde{f}^{\zeta^*} = \tilde{f}^{\zeta+\epsilon} = \tilde{k}_{\text{model}} \tilde{f}^{\epsilon}, \tag{A7}$$

where $\tilde{k}_{\text{model}}$ is the stiffness ratio predicted by the model. The choice of representing the epistemic error by $\sigma_\zeta^2$ (rather than $\sigma_{\tilde{k}}^2$) has the consequence of treating the errors in $\tilde{k}$ as multiplicative.

The average power-law exponent $\zeta$ from $M$ cantilevers can be estimated by

$$\zeta_{\text{est}} = \frac{1}{M} \sum_{m=1}^{M} \frac{\log \tilde{k}}{\log \tilde{f}}, \tag{A8}$$

which is equal to Eq. (3). Assuming that $\epsilon$ is independent of $\tilde{f}$, the variance $\sigma_\zeta^2$ can be estimated by

$$\sigma_{\zeta,\text{est}}^2 = \frac{1}{M-1} \sum_{m=1}^{M} \left[\frac{\log \tilde{k}}{\log \tilde{f}} - \zeta_{\text{est}}\right]^2. \tag{A9}$$

Now, the variance of $\tilde{k}$ can be related to $\sigma_\zeta^2$ by a first-order Taylor expansion

$$\sigma_{\tilde{k}} \approx \sigma_\zeta \frac{\partial \tilde{f}^\zeta}{\partial \epsilon} = \sigma_\zeta \tilde{f}^\zeta \log \tilde{f}, \tag{A10}$$

which is accurate in the limit of $\tilde{f}^{\sigma_\zeta} \cdot \log \tilde{f}^{\sigma_\zeta} \ll 1$. This equation is equal to Eq. (9) and used to as a measure of epistemic error when applying the frequency-ratio method for estimating $\tilde{k}$ from a measurement of $\tilde{f}$.

This entire analysis can be applied to separate eigenmodes by the substitution $\zeta \to \zeta_n$.

### c. In-factory error analysis of stiffness ratio

The measurement error in the stiffness ratio $\sigma_{k_n/k_1}$ during in-factory calibration combines contributions from the error in the laser spot versus tip correction $\sigma_x$, errors in the effective lever arms $\sigma_{r_n}$ and $\sigma_{r_1}$ from mode shape fitting, and random errors due to stochastic thermal noise for both normalized stiffness errors $\sigma_{k_n}$ and $\sigma_{k_1}$:

$$\sigma_{k_n/k_1}^2 = \frac{4}{L^2} \left[\frac{r_1 - r_n}{r_1 r_n}\right]^2 \sigma_x^2 + \frac{4\Delta x^2}{r_n^4 L^2} \sigma_{r_n}^2 + \frac{4\Delta x^2}{r_1^4 L^2} \sigma_{r_1}^2 \\ + \frac{\sigma_{k_1}^2}{k_1^2} + \frac{\sigma_{k_n}^2}{k_n^2}. \tag{A11}$$

Given the luxury of time during in-factory calibration, the thermal spectra were always averaged long enough to make the random stiffness errors $\sigma_{k_n}$ and $\sigma_{k_1}$ negligible compared to other errors. Furthermore, the contributions from stiffness errors from mode shape corrections $\sigma_{r_n}^2$ and $\sigma_{r_1}^2$ were minimized by keeping $\Delta x$ as close to 0 as possible[2].

Additionally, the frequency-response of the LDV may cause a repeatable absolute error in estimating $k_n/k_1$. The accuracy error $\epsilon_n$ (in dB) causes a multiplicative error of the $n^{\text{th}}$ mode amplitude by a factor $10^{\epsilon_n/10}$. This results in an accuracy error of the $\zeta_n$ estimate that can be calculated by

$$\epsilon_{\zeta_n} = \frac{|\epsilon_n + \epsilon_1|}{10 \log_{10} \tilde{f}}. \tag{A12}$$

which can be related to $\sigma_{k_n/k_1}^2$ through Eq. ().

### d. AC approach curves

An example dataset for AC approach curves performed with an AC240 cantilever on a silicon sample are presented in Figure 8. The slope of each linear fit was used as a measure of the (inverse) OBD sensitivity.

---

[2] For cantilevers with no tip setback, $\Delta x < 4$ μm could be achieved given the small laser spot diameter (~2.5 μm). For cantilevers with a tip setback, $\Delta x < 1$ μm was achieved by positioning the laser spot above the tip (the tip position was determined from a side-view photograph of the cantilever).

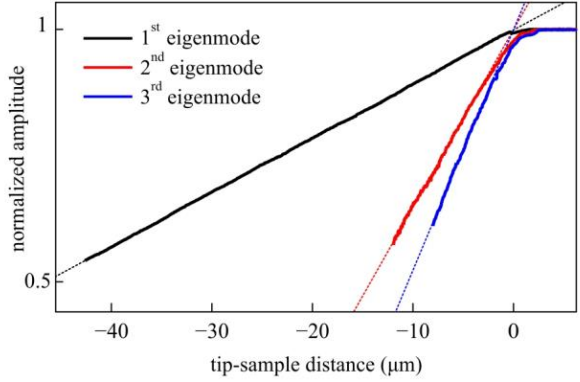

Figure 8: AC approach curves of the three first eigenmodes of an AC240 cantilever against a silicon sample, and corresponding linear fits.

### e. Benchmark study

The methodology described in this paper was benchmarked on a cantilever that closely resembles a perfect Euler-Bernoulli beam: a silicon cantilever (NOCAL, Bruker) with dimensions $395.7\,\mu m \times 28.4\,\mu m$, determined optically by a photograph acquired with the Cypher AFM (not shown).

The five first eigenmode shapes were mapped with LDV and are compared to the analytical Euler-Bernoulli model (not shown). Most of the disagreement between the mathematical model and LDV measurements was reconciled by FEA simulation of a NOCAL cantilever with 1) an infinitely rigid boundary at the base of the cantilever, and 2) a realistic silicon cantilever chip. While the FEA with an infinitely rigid boundary matched the mathematical Euler-Bernoulli model (as expected), the FEA with a realistic chip matched the LDV data, as seen in Figure 9. This concluded that modelling the cantilever chip is necessary for accurate FEA simulation of cantilever dynamics, and that this is especially true at increasing eigenmode numbers due to their increase in stiffness.

Secondly, the interferometric stiffness and resonance frequency measurements were compared to their FEA-simulated counterparts, as shown in Figure 10. The stiffness-to-frequency power law extracted from the FEA ($\zeta = 2.006 \pm 0.002$) and measured with the LDV ($\zeta = 2.002 \pm 0.003$) were close to the theoretical value.

This benchmark experiment sets a lower bound to the discrepancies expected between LDV measurements and FEA simulations and suggests it may be necessary to model the cantilever chip in these studies.

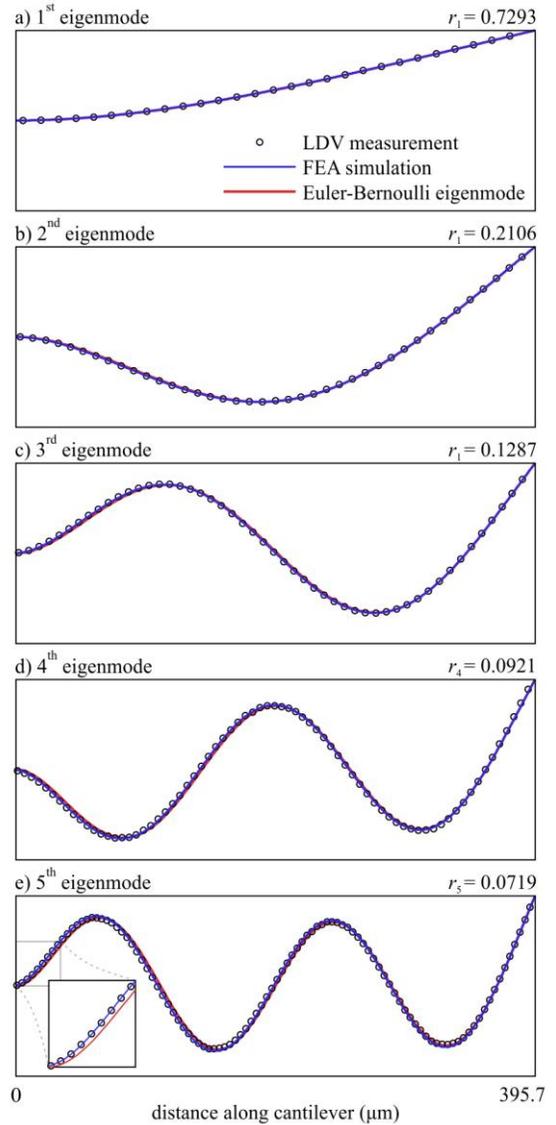

Figure 9: LDV measurements (piezoacoustically driven cantilever) and FEA simulations of the first five eigenmodes on a NOCAL cantilever, and the Euler-Bernoulli analytical eigenmodes. Including the cantilever chip in the FEA model was necessary achieve agreement between FEA and LDV, as clarified by the inset. The experimental lever arm ratios $r_n$ are listed for each mode. Note that the deflection of these driven modes (along the vertical axis) is highly exaggerated for visual reasons; the true deflections never exceeded 100nm.

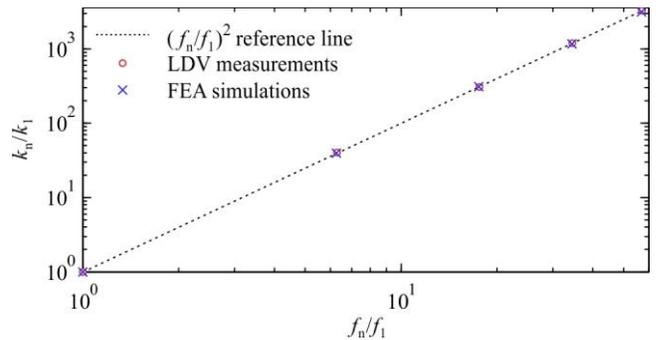

Figure 10: Power law relating the ratio of stiffnesses to the ratio of frequencies of the first five eigenmodes of a NOCAL cantilever, measured with three different methods.

## 10. Appendix B

This section presents four special cases of cantilever beams to provide an understanding of the prominent causes for deviation from ideal cantilever behavior, specifically deviations from the scaling law $k \propto \omega^2$. The goal is to assess the impact of each situation on the relationship between the stiffness and resonance frequency of different eigenmodes.

All four of these cases are solutions of the equation of motion for transverse free vibrations of a beam[67]

$$\frac{\partial^2}{\partial x^2}\left[\frac{EI(x)}{L^4}\frac{\partial^2 \Psi_n(x,t)}{\partial x^2}\right] + \rho_c A_c(x)\frac{\partial^2 \Psi_n(x,t)}{\partial t^2} = 0, \quad (B1)$$

where $A_c(x)$ is the cross-sectional area, $I(x)$ is the second moment of area, and $\rho_c$ is the mass density of the cantilever. By separation of variables, the eigenmode shape $\psi_n$ can be isolated from the time-dependent eigenfunction

$$\Psi_n(x,t) = \psi_n(x)\tau(t), \quad (B2)$$

leading to the simplified equation of motion:

$$\frac{\partial^2}{\partial x^2}\left[\frac{EI(x)}{L^4}\frac{\partial^2 \psi_n(x)}{\partial x^2}\right] - \omega_n^2 \rho_c A_c(x)\psi_n(x) = 0. \quad (B3)$$

### a. Special case: Euler-Bernoulli beam

An Euler-Bernoulli beam has a constant cross-sectional area $A_0$ and second moment of area $I_0$ along its entire length. Enforcing the boundary conditions

$$\psi(0) = \frac{d\psi}{dx}\bigg|_{x=0} = \frac{d^2\psi}{dx^2}\bigg|_{x=1} = \frac{d^3\psi}{dx^3}\bigg|_{x=1} = 0 \quad (B4)$$

results in the normalized solutions of the eigenmode equation

$$\psi_n(x) = \frac{[(\cos \kappa_n x - \cosh \kappa_n x) - K(\sin \kappa_n x - \sinh \kappa_n x)]}{[(\cos \kappa_n - \cosh \kappa_n) - K(\sin \kappa_n - \sinh \kappa_n)]}, \quad (B5)$$

where the Kappa-factor

$$K = \frac{\cos \kappa_n + \cosh \kappa_n}{\sin \kappa_n + \sinh \kappa_n}, \quad (B6)$$

and each modal wavenumber $\kappa_n$ is a root of the characteristic equation

$$1 - \cos \kappa x \cosh \kappa x = 0. \quad (B7)$$

This orthogonal basis of eigenfunctions $\psi_n(x)$ has corresponding angular resonance frequencies

$$\omega_n = \frac{\kappa_n^2}{L^2}\sqrt{\frac{EI_0}{\rho_c A_c}} \quad (B8)$$

and respective spring constants

$$k_n = \frac{\kappa_n^4}{4}\frac{EI_0}{L^3}. \quad (B9)$$

This leads to the relationship

$$k_n = \left(\frac{\rho_c A_c L}{4}\right)\omega_n^2 \quad (B10)$$

that results in the well-known scaling law

$$\frac{k_n}{k_1} = \left(\frac{\omega_n}{\omega_1}\right)^2, \quad (B11)$$

which is often used to estimate higher mode spring constants. This scaling is plotted in Figure 11 for the first five eigenmodes of an Euler-Bernoulli beam.

### b. Special case: uniform beam with tip mass

Adding a tip mass $m_{\text{tip}}$ at the very end of an Euler-Bernoulli beam can be modelled by updating the third-order boundary condition Eq. (B2) to

$$EI\frac{\partial^3 \Psi}{\partial x^3}\bigg|_{x=1} = m_{\text{tip}}\frac{\partial^2 \Psi}{\partial t^2}\bigg|_{x=1}. \quad (B12)$$

In this case, the added mass results in a decrease in each modal wavenumber $\kappa_n$. Interestingly, the same eigenmode from Eq. (B3) still applies; however, the modal wavenumbers are instead taken as roots from the modified characteristic equation that generalizes Eq. (B7) into

$$1 - \cos \kappa x \cosh \kappa x \left(1 + RL\kappa(\tan \kappa x - \tanh \kappa x)\right) = 0, \quad (B13)$$

where the mass ratio

$$R = \frac{m_{\text{tip}}}{\rho_c A_c L}. \quad (B14)$$

Equation (B8) can be used to calculate $\omega_n$ in this case, because the wave velocity is unaffected by the tip mass and constant throughout the full length of the beam. However, the stiffness must be calculated using Eq. (8), because the mode shape is affected by the added mass, leading to

$$k_n = \frac{EI_0}{L^3}\int_0^1 \left|\frac{\partial^2 \psi_n}{\partial x^2}\right|^2 dx. \quad (B15)$$

In this case, the squared scaling law of Eq. (B11) breaks down. This is shown in Figure 11, where the stiffness was plotted for $R = 0.1$. Applying the $\zeta$ power-law approximation leads to $\zeta > 2$ for $R > 0$.

### c. Special case: uniform beam with tip setback

For an Euler-Bernoulli with a massless tip positioned with a setback $\Delta x$ from the cantilever end, the effective stiffness eigenmodes increases without any consequence on the eigenmode frequencies. Given the boundary condition from Eq. (B4) that ensures no curvature of the cantilever at its end, a linear correction can be applied to calculate reduction in measured amplitude due to tip setback, as in

$$A_{\text{tip}} = A_{\text{end}}\left(1 + \frac{\Delta x_t}{L}\frac{\partial \psi_n}{\partial x}\right). \quad (B16)$$

Now, the stiffness $k_n$ at the cantilever tip can be calculated form the stiffness $k_{n,\text{end}}$ at the cantilever end by

$$k_n = k_{n,\text{end}} \left(1 + \frac{\Delta x_t}{L} \frac{\partial \psi_n}{\partial x}\right)^{-2}. \quad (B17)$$

Here, the lever arm ratio $r_n$ presented in Eq. (5) can be formally defined as

$$r_n = \left(\frac{\partial \psi_n}{\partial x}\right)^{-1}\bigg|_{x=1}. \quad (B18)$$

Note that the linear approximation of the cantilever end is only appropriate for $|\Delta x/L| < r_n/2$ (approximately).

For an Euler-Bernoulli beam, Eq. (B16) has an analytical solution. They were used to plot the first five stiffness ratio in Figure 11 for a tip setback $\Delta x/L = -0.03$. It can be shown that a power-law approximation for this effect would result in $\zeta > 2$ for $\Delta x/L < 0$.

### d. Special case: triangular cantilever

For a triangular tipless cantilever, $I(x)$ and $A(x)$ vary linearly from base to tip as

$$I(x) = I_0(1-x) \quad (B19)$$

and

$$A_c(x) = A_0(1-x). \quad (B20)$$

Solutions to Eq. (B3) under these conditions can be expressed as an infinite sum of Euler-Bernoulli eigenmodes, or approximated by a finite sum of $M$ eigenmodes, as in

$$\psi_n(x) = \sum_m^M c_m \psi_m^{\text{EB}}(x). \quad (B21)$$

where $\psi_m^{\text{EB}}(x)$ represents the Euler-Bernoulli eigenfunction. Because the boundary conditions are fulfilled by all $\psi_m^{\text{EB}}(x)$, the only requirement is to minimize the objective function

$$\Lambda(\omega, c_1, c_2, \dots) = \left\|\frac{\partial^2}{\partial x^2}\left[\frac{EI(x)}{L^4}\frac{\partial^2 w_n(x)}{\partial x^2}\right] - \omega^2 \rho_c A_c(x) \frac{\partial^2 \psi_n(x)}{\partial t^2}\right\|^2, \quad (B22)$$

where the $\|\dots\|$ represents the vector norm.

Conveniently, a triangular cantilever results in an analytical $\Lambda$ that can be minimized for any choice of $\omega$ by finding the optimal $c_n$'s through least-squares. Each local minimum in $\Lambda$ corresponds to an eigenmode frequency $\omega_n$.

This procedure was performed for the first five eigenmodes of a triangular cantilever (with a length much larger than base width). Both the stiffnesses and frequencies of higher eigenmodes drop significantly relative to the Euler-Bernoulli beam, as shown in Figure 11. Applying the $\zeta$ power-law approximation leads to $\zeta \sim 1.5$ in this case.

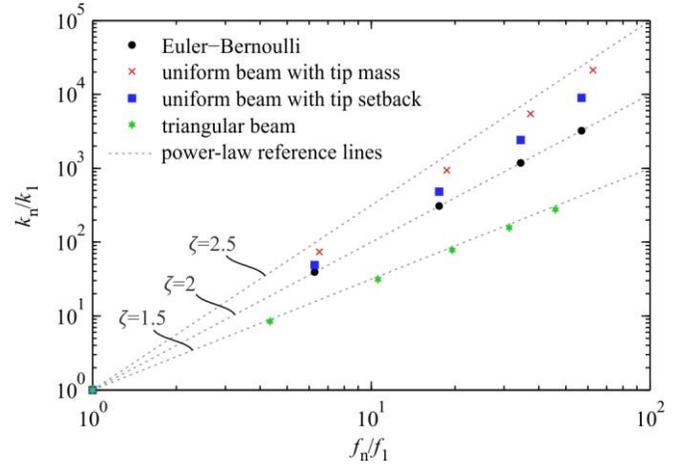

Figure 11: The analytically calculated changes in stiffness and resonance frequency of four special cases of cantilevers.